# Comprehensive Control of Optical Polarization Anisotropy in Semiconducting Nanowires

*Lei Fang[1], Xianwei Zhao[1], Yi-Hsin Chiu[1], Dongkyun Ko[1], Kongara M. Reddy[2], Nitin P. Padture[2], Fengyuan Yang[\*1] and Ezekiel Johnston-Halperin[\*1]*

ABSTRACT

The demonstration of strong photoluminescence polarization anisotropy in semiconducting nanowires embodies both technological promise and scientific challenge. Here we present progress on both fronts through the study of the photoluminescence polarization anisotropy of randomly oriented nanowire ensembles in materials without/with crystalline anisotropy, small/wide bandgap, and both III-V/II-VI chemistry (InP/ZnO nanowires, respectively). Comprehensive control of the polarization anisotropy is realized by dielectric matching with conformally deposited $Ta_2O_5$ (dielectric ratios of 9.6:4.41 and 4.0:4.41 for InP and ZnO, respectively). After dielectric matching, the polarization anisotropy of the nanowire ensembles is reduced by 86% for InP:$Ta_2O_5$ and 84% for ZnO:$Ta_2O_5$.

[1]*Department of Physics, The Ohio State University, Columbus, OH 43210-1117, USA*

[2]*Department of Materials Science and Engineering, The Ohio State University, Columbus, OH 43210, USA*

\*email: fyyang@mps.ohio-state.edu, ejh@mps.ohio-state.edu

Semiconducting nanowires have attracted significant attention due to the unique electric and optical properties arising from their quasi-1D geometry.[1-3] Among these properties, optical polarization anisotropy is of particular importance for fundamental studies of polarization-sensitive electronic states and potential applications in polarization-sensitive photodetectors.[4-6] Detailed control of this polarization anisotropy is critical for progress on both fronts and is the focus of this investigation. The polarization anisotropy for luminescence from a single nanowire is defined as $\rho = (I_{||} - I_{\perp})/(I_{||} + I_{\perp})$, where $I_{||}$ ($I_{\perp}$) is the intensity parallel (perpendicular) to the nanowires axis. The most general source of this anisotropy is the variation in dielectric constant between the nanowire ($\varepsilon$) and its environment ($\varepsilon_e$).[6-7] As a result, tuning $\varepsilon$ and $\varepsilon_e$ directly controls the optical polarization anisotropy. Initial attempts to implement this scheme have been demonstrated for individual chemically-synthesized CdSe nanostructures[8] Here we demonstrate a general approach to dielectric matching as applied to InP and ZnO nanowires. Conformal sputter coating of $Ta_2O_5$ results in an anisotropy reduction of 86% for InP:$Ta_2O_5$ and 84% for ZnO:$Ta_2O_5$ ($\varepsilon_{InP}$:$\varepsilon_{Ta2O5}$= 9.6:4.41 and $\varepsilon_{ZnO}$:$\varepsilon_{Ta2O5}$= 4.0:4.41).[9-10] This success validates a general scheme to control the optical polarization anisotropy in materials without/with crystalline anisotropy, with narrow/wide band gap, and with III-V/II-VI chemistry (InP and ZnO nanowires, respectively).

It has been predicted by theory[11-12] and observed experimentally[13-14] that in ensembles of randomly oriented nanowires there is a polarization memory effect.[11] Specifically, in nanowire ensembles the measured optical polarization anisotropy (which depends on both the absorption anisotropy and luminescence anisotropy) does not disappear, but rather decreases as compared to the single nanowire value due to weighted ensemble averaging. Figure 1a shows a schematic describing how this residual emission polarization can occur in nanowire ensembles. As a result, the ensemble studies presented here can be quantitatively correlated with single nanowire anisotropies. A detailed model for this correlation will be developed below, but first we address the synthesis and measurement of nanowire ensembles. InP and ZnO nanowires are grown using pulsed laser deposition (PLD)[15] and carbothermal reduction synthesis,[16] respectively. The substrate (a silicon wafer with 300 nm thermal oxide) is treated with diluted 50nm Au colloid as a growth catalyst. For PLD growth of InP nanowires, In and P vapors are generated by ablating a finely ground InP target using an excimer laser and carried by Ar gas to the preheated substrate (temperature 450°C to 550°C). For ZnO nanowires, a 1:1 molar mixture of ZnO and graphite powders is heated at the center of the furnace to 900 °C. The Zn vapor is then carried by a 2% $O_2$ in Ar gas flow to the substrate.[16] The nanowires are induced to lie flat by allowing a drop of methanol to evaporate from the substrate and dielectric matching is realized by coating the as-grown nanowires with a conformal $Ta_2O_5$ layer in an ultra high vacuum (UHV) off-axis sputtering system in an atmosphere of 5% $O_2$ in Ar with a total pressure of 18 mTorr.

Figures 1b and 1c show transmission electron microscopy (TEM) images for as-grown InP and ZnO nanowires, respectively. InP (ZnO) nanowires have a single crystalline core of ~ 50 nm (~ 30 nm) and an amorphous oxide shell formed during growth with a thickness of ~ 7 nm (~ 5 nm). The diffraction pattern in the core indicates that the nanowire has zincblende (ZB) structure with a <111> growth direction for InP nanowires (Fig. 1b, inset) and a wurzite (WZ) structure with a <001> growth direction for ZnO nanowires (Fig. 1c, inset). Previous studies[17-18] have shown that this growth direction for ZnO results in a significant dipole moment perpendicular to the nanowire axis, leading to a significant crystal-symmetry driven polarization anisotropy for an individual ZnO nanowire ($\rho$ from -0.5 to -0.65). Here the negative sign of $\rho$ indicates that $I_\perp$ is more intense than $I_\parallel$, as expected when the crystal-symmetry driven polarization anisotropy dominates the dielectric anisotropy.

For photoluminescence measurements, a wavelength of 690 nm (345 nm) and power of 2 mW (0.2 mW) is focused to a spot size of ~100 um to excite luminescence from the InP (ZnO) nanowire ensemble. To measure the polarization anisotropy a Glan-Laser linear polarizer is placed in the pump path and a second linear polarizer and an achromatic half-waveplate are placed in the collection path to allow measurement of the polarization of the nanowire luminescence. Spectral resolution is provided by a 0.3 m spectrometer and a liquid $N_2$ cooled CCD camera. The ensembles are maintained at a temperature of 5 K to maximize PL intensity.

The inset in Fig. 2a shows a scanning electron microscopy (SEM) image of as-grown InP nanowires. The nanowires are randomly oriented with a typical diameter of 50 – 80 nm and length longer than 10 um. A characteristic spectrum for the as-grown ensemble is plotted in Fig. 2a, revealing a peak in nanowire luminescence centered at 849 nm. The luminescence intensity is calculated by integrating the intensity through the shaded region (830 nm to 878 nm). Since there is no uniquely defined nanowire axis for nanowire ensembles, $\langle I \rangle_\parallel$ ($\langle I \rangle_\perp$) is defined as the ensemble luminescence intensity when the polarization of excitation is parallel (perpendicular) to the luminescence (Fig. 2a; black and red spectra, respectively). The polarization anisotropy for this nanowire ensemble is then $\langle \rho \rangle = (\langle I \rangle_\parallel - \langle I \rangle_\perp)/(\langle I \rangle_\parallel + \langle I \rangle_\perp) = 0.25$. The inset in Fig. 2b shows an SEM image of the nanowires after they have been conformally coated with an average of 670 ± 34nm of Ta$_2$O$_5$. The polarization resolved photoluminescence spectrum (Fig. 2b) reveals that the intensity is nearly identical for perpendicular and parallel configurations leading to $\langle\rho\rangle \sim 0.036$, a reduction of 86%, consistent with the qualitative expectation that the dielectric matching will suppress the polarization anisotropy.

ZnO nanowires, as can be seen in Fig. 2c, range from 30 nm to 50 nm in diameter and exhibit spectrally sharp and bright photoluminescence centered at 368.5 nm. The polarization anisotropy is found to be $\langle\rho\rangle = 0.19$. However, as mentioned above, this anisotropy has a second potential origin: the presence of a strong crystalline anisotropy perpendicular to the nanowire axis.[19-20] The ensemble measurements presented here do not distinguish between orientations parallel and perpendicular to the nanowire axis. As a result both dielectric and crystalline anisotropy have the same symmetry for these measurements. After a conformal coating of 235 ± 3nm of Ta$_2$O$_5$, ensemble anisotropy measurements (Fig. 2d) show that the polarization anisotropy decreases to $\langle\rho\rangle = 0.031$ (an 84% suppression). This result suggests that while crystalline anisotropy may be present in these ZnO nanowires, it must be relatively weak and the dielectric anisotropy plays a dominant role.

In order to gain a quantitative understanding of our results we construct a simple model of our ensembles,[21] shown schematically in inset of Fig. 3a. We assume that the nanowires lie in the plane of the substrate with random in-plane orientation. For an individual nanowire, the angle between the nanowire axis and the excitation field $E$ is $\varphi$ and the angle between the excitation and detection is $\theta$. The total luminescence intensity of the ensemble then corresponds to the integration over $\varphi$ from 0° to 360°, which yields the following expression for the total ensemble intensity:

$$\langle I \rangle = \frac{\pi E^2 L^2 \left[(10\varepsilon_e^2 + 4\varepsilon_e\varepsilon + 2\varepsilon^2)(9\varepsilon_e^2 + 2\varepsilon_e\varepsilon + \varepsilon^2) + (\varepsilon_e - \varepsilon)^2(3\varepsilon_e + \varepsilon)^2 \cos 2\theta\right]}{4(\varepsilon_e + \varepsilon)^2(3\varepsilon_e^2 + 2\varepsilon_e\varepsilon + \varepsilon^2)}$$

…..(1).

This parameter-free model describes the dependence of the ensemble luminescence on the angle between the excitation and detection polarization for nanowire ensembles in any dielectric environment.

Figure 3a shows the full angular dependence of the normalized PL intensity in as-grown (red dots) and dielectric-matched (black triangles) InP nanowires. The PL intensity shows a clear $\cos(2\theta)$ dependence and confirms that the suppression in anisotropy (from $\langle\rho\rangle = 0.25$ to 0.036) extends for all values of $\theta$. The curves calculated from Eq. (2) are plotted as blue and green lines for bare and coated nanowires, respectively. Given that no free parameters are used in this calculation, the experimental and calculated curves are in good agreement; A plausible explanation is that the polycrystalline debris produced by the PLD process generates contaminating luminescence that does not exhibit significant anisotropy and cannot be spectrally distinguished from the nanowire luminescence.

Similarly for ZnO nanowires (Fig. 3b), the dielectric matching suppresses the polarization anisotropy from $\langle\rho\rangle = 0.19$ to 0.031. A small deviation from the calculated value for the as-grown nanowires is also observed, likely due to similar spectral contamination from undesirable growth modes. The dielectric model predicts a polarization anisotropy of $\langle\rho\rangle < 0.004$ for ZnO:Ta$_2$O$_5$ (the inequality arises from variation of $\varepsilon_{Ta2O5}$ in the literature[10]). The observed anisotropy ($\langle\rho\rangle = 0.031$ (Fig. 3b black triangles), coupled with the fact that this ensemble measurement does not distinguish between axial and radial anisotropy, suggests that a residual crystalline anisotropy may be present. Applying an analysis similar to that developed for the dielectric anisotropy yields a single-wire anisotropy[21] of $\rho$ = -0.27, significantly smaller than the value reported in the literature[17-18] ($\rho$ from -0.5 to -0.65). While a detailed exploration of this crystalline anisotropy is beyond the scope of this study, a possible explanation may lie in the significantly smaller diameter of the ZnO wires used in this study (30 – 50 nm versus 150 – 500 nm, respectively) giving rise to a strain-induced variation in the c-axis lattice constant.[17-18]

In conclusion, a general approach to controlling the polarization anisotropy in semiconducting nanowires is demonstrated. Significant suppression of this anisotropy is demonstrated by dielectric matching the nanowires with a conformal coating of $Ta_2O_5$. While a residual crystalline anisotropy is found in ZnO wires, it is clear that the strategy of dielectric matching to control optical polarization anisotropy applies across a wide class of nanowire materials. This work opens the door to both fundamental studies of quasi-1d systems and applications such as multi-spectral polarization-sensitive photodetectors and polarization-controlled light-emitting diodes.

**Acknowledgements**

The authors gratefully acknowledge financial support from DOE grant number DE-SC0001304 for this research. Additional support was provided by NSF through the MRSEC program under grant number DMR-0820414. We also thank Dr. D. Li for TEM assistance.

**Figure Captions**

**Figure 1.** (a) Linearly polarized light preferentially excites nanowires oriented parallel to the polarization and consequently luminescence with the same linear polarization is emitted from those nanowires. (b) TEM image of an as-grown InP nanowire with single crystalline core (50 nm) – amorphous shell (7 nm) structure. Inset: Diffraction pattern showing a zincblende structure and <111> growth direction. (c) TEM image of an as-grown ZnO nanowire, with single crystalline core (30nm) – amorphous shell (5nm) structure. Inset: Diffraction pattern showing a wurtzite structure and<001> growth direction.

**Figure 2.** (a), (b) Photoluminescence for as-grown and oxide coated InP nanowires, respectively. Inset: corresponding SEM images. (c), (d) Photoluminescence spectra for as-grown and $Ta_2O_5$ coated ZnO nanowires, respectively. Inset: corresponding SEM images. The black (red) curves are for excitation polarization parallel (perpendicular) to the detection polarization.

**Figure 3.** (a) Comparison between the experimental (red dots and black triangles) and calculated (blue and green lines, see Eq. 2 in the main text) normalized PL intensity for as-grown and $Ta_2O_5$ coated InP nanowire ensembles, respectively. Inset: Schematic representation of the model presented in Eq. 1. (b) The same comparison for as-grown and $Ta_2O_5$ coated ZnO nanowires.

Figure 1

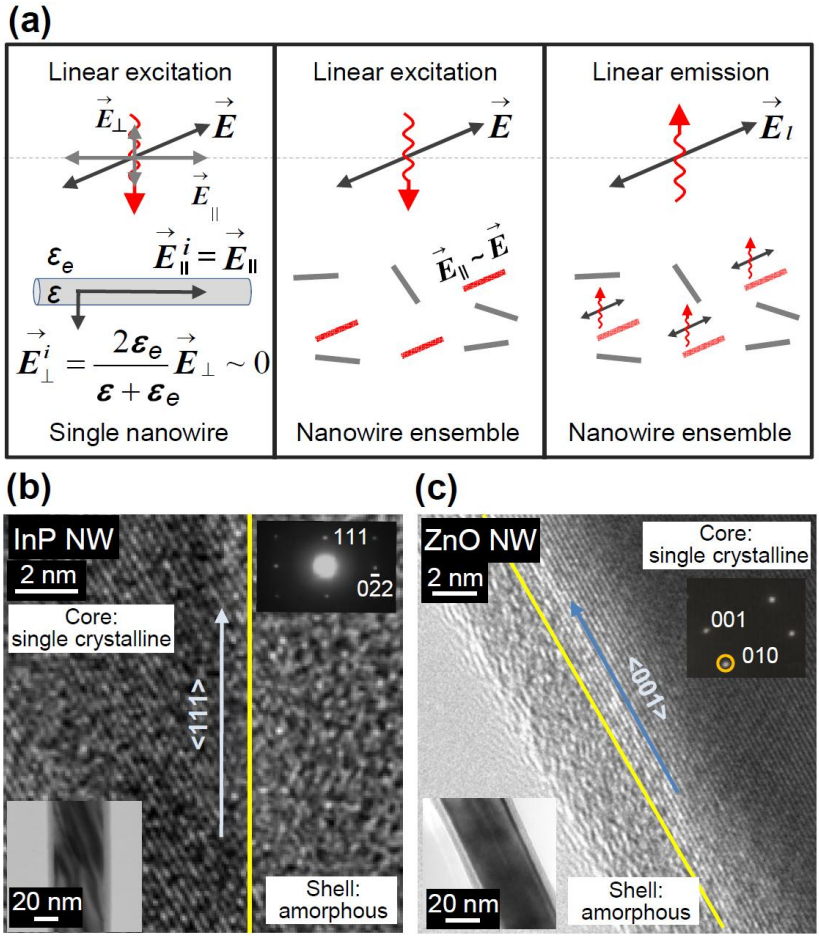

Figure 2

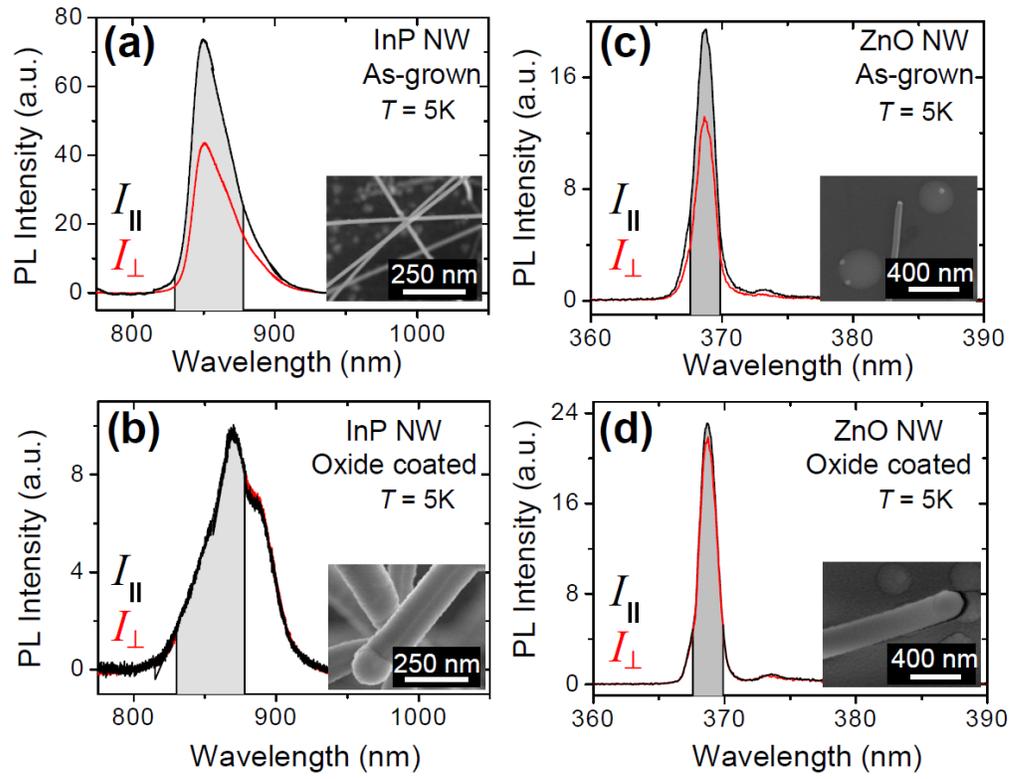

Figure 3

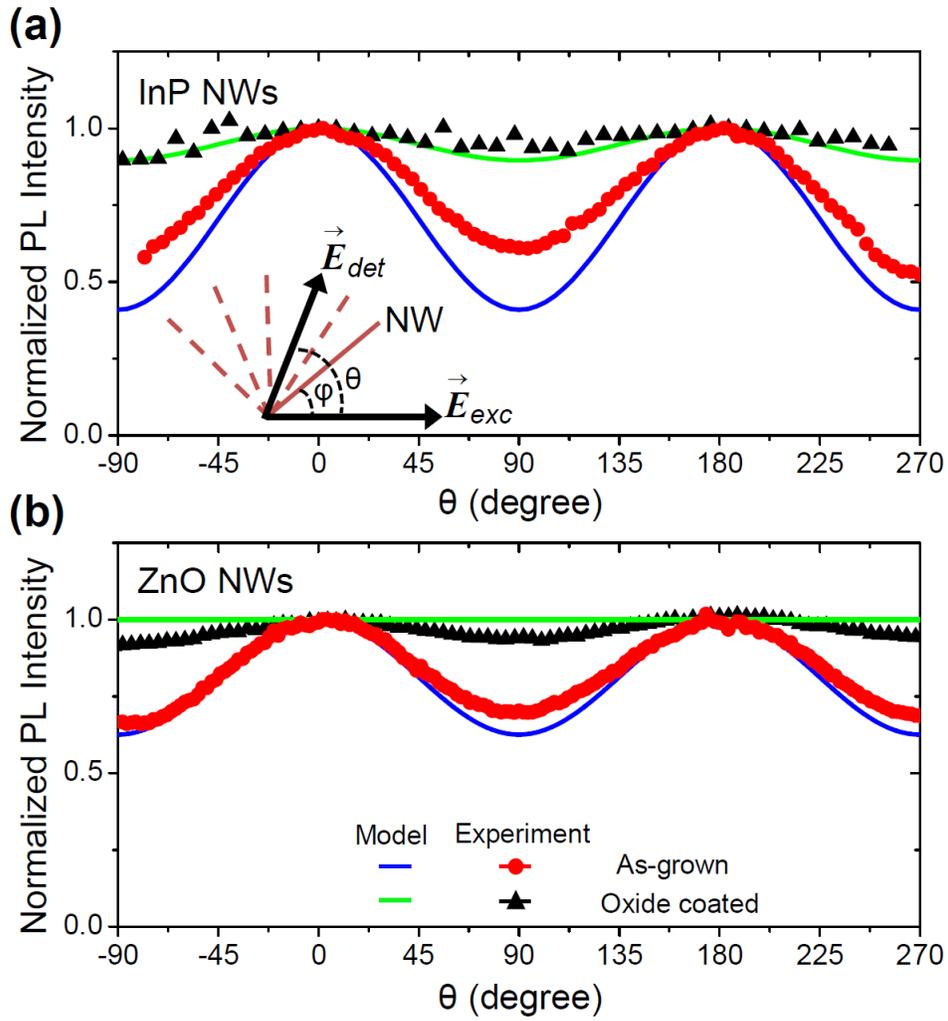

Supporting information

Discussion S1:

As mentioned in main text, we assume that the nanowires lie in the plane of the substrate with random in-plane orientation. For an individual nanowire, the angle between the nanowire axis and the excitation field $E$ is $\varphi$ and the angle between the excitation and detection is $\theta$. The detected luminescence intensity from the single wire can therefore be written as:

$$I = (E\cos\varphi)^2 \times L^2[\cos^2(\theta-\varphi) + \frac{6\varepsilon_e^2}{(\varepsilon+\varepsilon_e)^2 + 2\varepsilon_e^2}\sin^2(\theta-\varphi)]$$

$$+ (\frac{2\varepsilon_e}{\varepsilon+\varepsilon_e}E)^2(\sin\varphi)^2 \times L^2[\cos^2(\theta-\varphi) + \frac{6\varepsilon_e^2}{(\varepsilon+\varepsilon_e)^2 + 2\varepsilon_e^2}\sin^2(\theta-\varphi)]$$

…………………..(1)

where the anisotropy of the optical absorption is[S1] $\frac{\alpha_\perp}{\alpha_\parallel} = \left(\frac{E_{exc}^\perp}{E_{exc}^\parallel}\right)^2 = \left(\frac{2\varepsilon_e}{\varepsilon+\varepsilon_e}\right)^2$ and the anisotropy of the light emission[S2] is $\frac{I_\perp}{I_\parallel} = \frac{6\varepsilon_e^2}{(\varepsilon+\varepsilon_e)^2 + 2\varepsilon_e^2}$. Here $I_\parallel$ ($I_\perp$) is the luminescence intensity of the individual nanowire when $\varphi$ is 0° (90°). $E^2$ is the intensity of excitation and $L^2$ is a dimensionless factor that represents the intensity of emission. The total luminescence intensity of the ensemble includes nanowires of all orientations, corresponding to the integration over $\varphi$ from 0° to 360°, which yields the following expression for the total ensemble intensity:

$$\langle I \rangle = \frac{\pi E^2 L^2 \left[(10\varepsilon_e^2 + 4\varepsilon_e\varepsilon + 2\varepsilon^2)(9\varepsilon_e^2 + 2\varepsilon_e\varepsilon + \varepsilon^2) + (\varepsilon_e - \varepsilon)^2(3\varepsilon_e + \varepsilon)^2 \cos 2\theta\right]}{4(\varepsilon_e + \varepsilon)^2(3\varepsilon_e^2 + 2\varepsilon_e\varepsilon + \varepsilon^2)}$$

…..(2)

This parameter-free model describes the dependence of the ensemble luminescence on the angle between the excitation and detection polarization for nanowire ensembles in any dielectric environment.

Discussion S2:

For nanowire ensembles, the crystalline anisotropy can be modeled by the same approach as the dielectric anisotropy. With the same geometry discussed in the main text, the detected luminescence intensity from a single wire can be written as:

$$I = (E\cos\varphi)^2 \times L^2[\cos^2(\theta-\varphi) + C\sin^2(\theta-\varphi)]$$
$$+ C(E\sin\varphi)^2 \times L^2[\cos^2(\theta-\varphi) + C\sin^2(\theta-\varphi)]$$

……..(S1)

Here we assume the crystalline anisotropy for the optical absorption and light emission is the same and it is defined as $\frac{\alpha_\perp}{\alpha_\parallel} = \frac{I_\perp}{I_\parallel} = C$. The total intensity with only crystalline anisotropy can then be obtained by the integration over the angle $\varphi$ from 0 to 360 degrees：

$$\langle I \rangle = \frac{\pi E^2 L^2 \left[2(1+C)^2 + (C-1)^2 \cos 2\theta\right]}{4}$$

……..(S2)

Using the normalized experimental value $\frac{\langle I \rangle_{(\theta=\frac{\pi}{2})}}{\langle I \rangle_{(\theta=0)}} = 0.93$ (which gives $\langle \rho \rangle = 0.031$) for the Ta$_2$O$_5$ coated ZnO nanowires, C is calculated to be $\frac{I_\perp}{I_\parallel} = 1.737$, which means the corresponding crystalline anisotropy for a single nanowire is $\rho = (I_\parallel - I_\perp)/(I_\parallel + I_\perp) = -0.27$.